\documentclass[reprint,superscriptaddress,showpacs,aps,prx]{revtex4-2}
\usepackage[pdftex]{graphicx}
\usepackage{amsmath}
\usepackage{amssymb}
\usepackage{placeins}
\usepackage{color}
\usepackage{xfrac}
\usepackage{hyperref}
\usepackage{xcolor}
\usepackage{siunitx}
\usepackage{enumitem}
\usepackage{xspace}
\hypersetup{
    colorlinks,
    linkcolor={red!50!black},
    citecolor={blue!50!black},
    urlcolor={blue!80!black}
}

\newcommand{\mps}{Mn$_\mathrm{1.4}$PtSn\xspace}

\providecommand{\vect}[1]{{\boldsymbol{#1}}}
\DeclareUnicodeCharacter{2212}{-}

\newcommand{\MPI}{Max Planck Institute for
Chemical Physics of Solids, 01187 Dresden, Germany.}
\newcommand{\DCN}{Dresden Center for Nanoanalysis, cfaed, TUD University of Technology Dresden, 01069 Dresden, Germany.} 
\newcommand{\IFW}{ Leibniz Institute for Solid State
and Materials Research Dresden, 01069 Dresden, Germany.}
\newcommand{\IFMP}{Institute for Solid State and Materials Physics, TUD University of Technology Dresden, 01062 Dresden, Germany.}
\newcommand{\DiamondLight}{Diamond Light Source, Harwell Science and Innovation Campus, Didcot, OX11~0DE, United Kingdom}
\newcommand{\Clarendon}{Department of Physics, Clarendon Laboratory, University of Oxford, Oxford, OX1~3PU, United Kingdom}
\newcommand{\ALBA}{ALBA Synchrotron Light Source, E-08290 Cerdanyola del Vallés, Barcelona Spain}

\newcommand{\Bessy}{Helmholtz-Zentrum Berlin für Materialien und Energie, D-12489 Berlin, Germany}
\newcommand{\HZDR}{Dresden High Magnetic Field Laboratory (HLD-EMFL), Helmholtz-Zentrum Dresden-Rossendorf, 01328 Dresden, Germany.}
\newcommand{\UNIDue}{Faculty of Physics and Center for Nanointegration Duisburg-Essen (CENIDE), University of Duisburg-Essen, 47057 Duisburg, Germany}
\newcommand{\ctqmat}{W{\"u}rzburg-Dresden Cluster of Excellence ct.qmat, TUD University of Technology Dresden, 01062 Dresden, Germany}
\newcommand{\uniA}{Experimental Physics IV, Center for Electronic Correlations and Magnetism, University of Augsburg, 86159 Augsburg, Germany}

\begin{document}


\title{Field-induced condensation of $\pi$ to $2\pi$ soliton lattices in chiral magnets}

\author{M. Winter}
\affiliation{\MPI}
\affiliation{\DCN}
\affiliation{\IFW}

\author{A. Pignedoli}
\affiliation{\UNIDue}

\author{M. C. Rahn}
\affiliation{\uniA}
\affiliation{\IFMP}

\author{A. S. Sukhanov}
\affiliation{\uniA}
\affiliation{\IFMP}

\author{B. Achinuq}
\affiliation{\Clarendon}

\author{J. R. Bollard}
\affiliation{\Clarendon}
\affiliation{\DiamondLight}

\author{M. Azhar}
\affiliation{\UNIDue}

\author{\mbox{K. Everschor-Sitte}}
\affiliation{\UNIDue}

\author{D. Pohl}
\affiliation{\DCN}

\author{S. Schneider}
\affiliation{\DCN}

\author{A. Tahn}
\affiliation{\DCN}

\author{V. Ukleev}    
\affiliation{\Bessy}

\author{M. Valvidares}
\affiliation{\ALBA}

\author{A. Thomas}
\affiliation{\IFW}
\affiliation{\IFMP}

\author{D. Wolf}
\affiliation{\IFW}

\author{\mbox{P. Vir}}
\affiliation{\MPI}

\author{T. Helm}
\affiliation{\HZDR}

\author{G. van der Laan}
\affiliation{\DiamondLight}

\author{T. Hesjedal}
\affiliation{\Clarendon}
\affiliation{\DiamondLight} 

\author{J. Geck}
\affiliation{\IFMP}
\affiliation{\ctqmat}

\author{C. Felser}
\affiliation{\MPI}
\affiliation{\ctqmat}

\author{B. Rellinghaus}
\affiliation{\DCN}

\date{\today}

\begin{abstract}
Chiral soliton lattices (CSLs) are nontrivial spin textures that emerge from the competition between Dzyaloshinskii–Moriya interaction, anisotropy, and magnetic fields. 
While well established in monoaxial helimagnets, their role in materials with anisotropic, direction-dependent chirality remains poorly understood. 
Here, we report the direct observation of a tunable transition from $\pi$ to $2\pi$ soliton lattices in the non-centrosymmetric Heusler compound Mn$_\mathrm{1.4}$PtSn. 
Using Lorentz transmission electron microscopy, resonant elastic X-ray scattering, and micromagnetic simulations, we identify a $\pi$-CSL as the magnetic ground state, in contrast to the expected helical phase, which evolves into a classical $2\pi$-CSL under increasing out-of-plane magnetic fields. 
This transition is governed by a delicate interplay between uniaxial magnetocrystalline anisotropy and magnetostatic interactions, as captured by a double sine-Gordon model. 
Our analysis not only reveals the microscopic mechanisms stabilizing these soliton lattices but also demonstrates their general relevance to materials with $D_{2d}$, $S_4$, $C_{nv}$, or $C_n$ symmetries.
The results establish a broadly applicable framework for understanding magnetic phase diagrams in chiral systems, with implications for soliton-based spintronic devices and topological transport phenomena.
\end{abstract}

\maketitle

\section{Introduction}

Topologically nontrivial spin textures, such as (anti)skyrmions and chiral soliton lattices (CSLs), arise in magnetic systems where symmetric exchange, the asymmetric Dzyaloshinskii–Moriya interaction (DMI), and anisotropies compete under external magnetic fields. 
In most known cases, such as monoaxial chiral helimagnets like Cr$_{1/3}$NbS$_2$, Cr$_{1/3}$TaS$_2$, or Yb(Ni$_{1-x}$Cu$_x$)$_3$Al$_9$, CSLs emerge from a helical ground state stabilized by isotropic DMI and easy-plane anisotropy~\cite{Togawa2012, Zhang2021, Matsumura2017}. 
There, the soliton wave vector aligns along a single hard axis, and the lattice evolves continuously under field.
Recently, increased attention has turned to magnets with anisotropic DMI, particularly in non-centrosymmetric systems such as Mn$_\mathrm{1.4}$PtSn~\cite{Nayak2017}, which host a rich variety of magnetic textures including antiskyrmions~\cite{Peng2020,Ma2020,Jena2020,Back2020}. 
In these systems, the DMI favors opposite chirality along orthogonal directions in the crystallographic $ab$ plane, enabling nontrivial two-dimensional textures with strong shape dependence~\cite{Winter2022,ZunigaCespedes2021,Sukhanov2020}. 
However, the microscopic mechanisms governing field-driven transitions between these textures, and the question whether CSL-like states can arise from distinct ground states, remain poorly understood.

The relevance of soliton lattices extends well beyond magnetic systems. 
Analogous periodic topological structures occur in the physics of elementary particles, most notably in quantum chromodynamics~\cite{Liu2019,Eto2025,Amari2025}, where they describe modulated chiral condensates, and in nematic liquid crystals, where related field-theoretical models capture defect lattices and director modulations~\cite{Shen2022,Abdullah2022}. 
As in magnetic systems, field-driven transitions between such lattices, including the formation of topologically protected skyrmionic structures, play a central role in the underlying physics.

In the present work, we focus on the emergence of CSLs in \mps, a compound with $D_{2d}$ symmetry and anisotropic DMI. 
While prior work has largely emphasized skyrmionic textures, we demonstrate that the magnetic ground state of \mps\ is not a helix, but rather a stripe domain structure composed of alternating domains separated by $\pi$ Bloch walls, effectively a $\pi$-CSL. 
Upon increasing the out-of-plane magnetic field, this ground state continuously evolves into a conventional CSL with $2\pi$ solitons.
To uncover this transition, we combine Lorentz transmission electron microscopy (LTEM), resonant elastic X-ray scattering (REXS), and micromagnetic simulations. 
The field-tunable evolution of modulation length and harmonic content reveals a soliton condensation process governed by a delicate interplay of magnetocrystalline anisotropy and magnetostatic interactions. 
Using a double sine-Gordon model, we construct a magnetic phase diagram that explains the stability of both $\pi$- and $2\pi$-CSLs.
Importantly, this behavior is not unique to \mps. 
Our framework applies broadly to systems with anisotropic DMI and uniaxial anisotropy, including materials with $D_{2d}$, $S_4$, $C_{nv}$, and $C_n$ point group symmetries~\cite{Bogdanov1994}, as well as thin films of cubic chiral magnets under strain. 
This establishes a generalized route to soliton lattice formation and control in a wide range of magnetic materials, with implications for spintronic functionality, topological transport, and related fields of physics.

\section{Results and Discussion}
\subsection{Experimental Approach}
\begin{figure}[t]
    \includegraphics[width=0.46\textwidth]{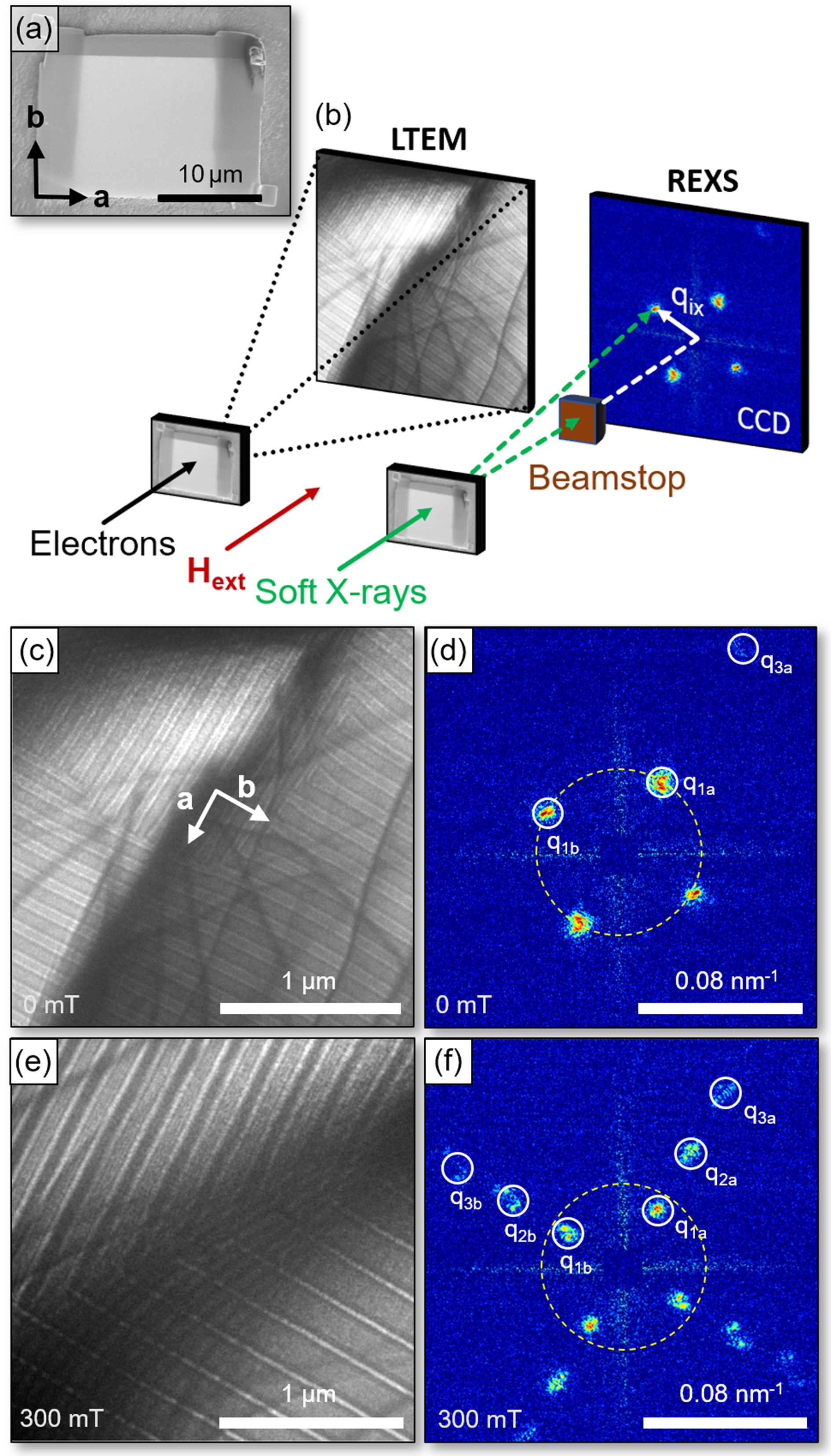}
    \caption{Experimental setup and basic experimental findings. 
    (a) SEM image of the \mps~lamella mounted across the square hole of a Mo carrier disc. 
    (b) Geometries of the LTEM and REXS experiments performed in transmission on the very same sample with magnetic fields applied along the direction of the impinging electron and X-ray beam, respectively. LTEM images and REXS patterns acquired at zero field and \SI{300}{\milli\tesla} are shown in (c-f). 
    The LTEM images in (c) and (e) were acquired at an overfocus of $288\,\mu$m. They reveal two perpendicularly oriented domains with stripy contrasts that alternate periodically along the crystallographic $a$ and $b$ axes of the \mps~lamella. The intensity of the background-subtracted REXS patterns in (d) and (f) is presented on a logarithmic color scale. The dashed yellow circle indicates the inverse modulation length in zero field, $q(\mu_0H_\mathrm{ext} = 0)$.}
    \label{Fig:LTEM_REXS_Exp}
\end{figure}

High-quality single crystals of \mps~ were grown by the self-flux method, as previously reported in Ref.~\cite{VirChemMat}. 
A thin lamella $(\SI{19}{\mu\meter}\times\SI{14.7}{\mu\meter}\times\SI{80}{\nano\meter})$ with a $c$-axis surface normal was cut from an oriented single crystal by focused ion beam (FIB) milling. 
As illustrated in Figs.~\ref{Fig:LTEM_REXS_Exp}(a,b), our experimental strategy is based on complementary insights in real and momentum space.
Thickness and shape variations of FIB-prepared lamellae make it crucial to perform these measurements on the very same specimen. 
To achieve this, the lamella was mounted on top of an X-ray opaque molybdenum disc (\SI{3}{mm} diameter, \SI{50}{\micro\meter} thickness) with an $8\times \SI{8}{\micro\meter}^2$ quadratic aperture that fits both commercial TEM holders and a dedicated sample carrier designed for our REXS experiments (see Methods section).

Figure \ref{Fig:LTEM_REXS_Exp}(c) shows the LTEM contrast of the magnetic ground state of \mps\ in zero magnetic field at room temperature. 
The image reveals two domains with stripe-type features that alternate periodically along the crystallographic $a$ and $b$ axes~\cite{Nayak2017,Ma2020,Peng2020}. 
In the corresponding REXS pattern in Fig.\ \ref{Fig:LTEM_REXS_Exp}(d), the magnetic scattering originating from the two perpendicular domain variants is combined to form a pseudo four-fold magnetic diffraction pattern. 
As indicated in the figure, the observed Bragg peaks are labeled ${q}_{{i,\xi}}$, where $i$ indicates the $i^\textrm{th}$ harmonic of the periodic stripe pattern and $\xi$ refers to domains whose stripes alternate along either the $a$ or $b$ crystallographic axis of the sample.
As shown in the figure, the Bragg peaks are labeled ${q}_{i,\xi}$, where $i$ denotes the $i^\mathrm{th}$ harmonic of the periodic stripe pattern, and $\xi$ indicates domains where the stripes alternate along the crystallographic $a$ or $b$ axes of the sample.

Upon applying an out-of-plane magnetic field along the crystallographic $c$ axis ($\vect{H}_\mathrm{ext}\parallel c$) as exemplified in Fig.\ \ref{Fig:LTEM_REXS_Exp}(e) for a field of ${H}_\mathrm{ext} = \SI{300}{\milli\tesla}$, the LTEM patterns change significantly: 
(i) The modulation length $L_\text{LTEM}(H_{\mathrm{ext}})$ is increased with respect to the patterns observed in zero field. 
(ii) The contrast modulation becomes asymmetric. Enlarged regions of homogeneous contrast are bordered by narrowed regions of alternating contrast. This indicates a broadening of stripes with a constant in-plane component of the magnetization like in field-polarized (i.e., out-of-plane magnetized) regions that are separated by narrowed stripes of modulated (in-plane) magnetization such as in domain walls (DWs).

These findings are corroborated by the results of our complementary REXS experiments. 
As shown in Figs.~\ref{Fig:LTEM_REXS_Exp}(d,f), the magnitude of the magnetic scattering vector $q_{{i,x}}(H_{\text{ext}})=2\pi/L_\text{REXS}(H_{\text{ext}})$ is reduced upon increasing the field. 
The field-induced occurrence of additional harmonics, i.e., the $\mathrm{2^{nd}}$ harmonics $q_{{2,a}}$ and $q_{{2,b}}$, and their field-dependent intensities reflect the asymmetric change of the modulation of the underlying magnetic structures already observed in the LTEM images.

\subsection{Analyzing the nature of the chiral soliton lattices from real space magnetic imaging}

\begin{figure*}[t]
    \includegraphics[width=\textwidth]{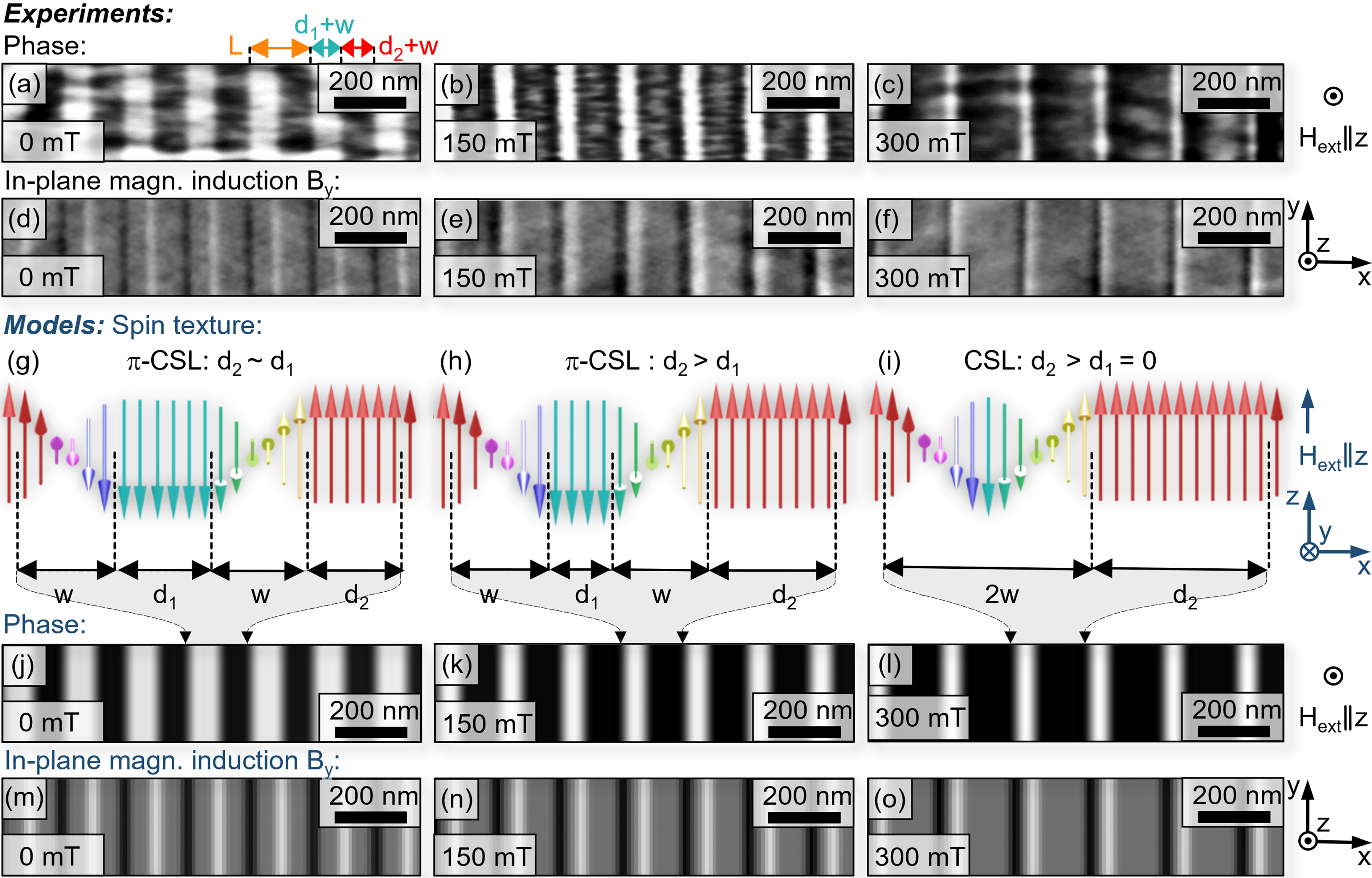}
    \caption{Analysis of the stripe contrast observed by LTEM in out-of-plane magnetic fields, $\mu_0 H$, of 0, 150, and \SI{300}{mT}.
    (a-c) Maps of the magnetic phase shift imposed on the electron wave as derived from TIE analyses of the LTEM images for sections of the stripe patterns. For this representation, all phase patterns were rotated within the $xy$ viewing plane to align the direction of alternating contrast with the abscissa, which is identified as $x$ direction; accordingly, the magnetic field is applied along the $z$ direction. Colored arrows indicate the periodic length, $L_\text{LTEM}$, and the widths of bright and dark stripe domains, $d_1$ and $d_2$, respectively (each including the domain wall width, $w$).
    (d-f) Maps of the $y$ component of the in-plane magnetic induction, $B_{y}$, derived from the phase maps using Eq.~\eqref{Eqn_TIE}.
    (g-i) Spin textures deduced from the induction maps and from the measures derived from phase maps assuming DWs with right-handed chirality. See text for details. 
    Corresponding phase and $B_{y}$ maps as calculated from the model spin textures in (g-i) are shown in panels (j-l) and (m-o), respectively.
    All phase and induction patterns are qualitative images that are contrast-enhanced for better visibility.}
    \label{Fig:LTEM_PhaseAnalysis}
\end{figure*}

In order to better understand the magnetic textures underlying the observed stripe patterns and their field-induced modification, we first analyze the LTEM images in more detail. 
The transport of intensity equation (TIE)~\cite{Teague(1983),Zuo2020,Lubk2024} is used to calculate the phase $\Phi$ of the electron wave from the intensity distribution in the LTEM images. 
Then, if the electrostatic potential of the sample is constant across the considered area, variations of the projected in-plane magnetic induction $\Vec{B}$ can be determined from this phase according to
        \begin{equation}
        \begin{split}
            \partial_x \Phi(\Vec{r}) = \frac{e}{\hbar} \int B_y(\Vec{r}_{\perp},z) dz = \frac{e}{\hbar} \bar{B}_y t \, , \\
            \partial_y \Phi(\Vec{r}) = - \frac{e}{\hbar} \int B_x(\Vec{r}_{\perp},z) dz = - \frac{e}{\hbar} \bar{B}_x t \, .
        \end{split}
  \label{Eqn_TIE}
  \end{equation} 
Here, $e$, $\hbar$, $\bar{B}_x$, $\bar{B}_y$, and $t$ denote the electron charge, Planck's constant, and the $x$ and $y$ components of the projected magnetic induction along $z$ throughout the thickness of the sample, respectively. The latter was determined right after the ion beam cutting in the FIB to be $t \lesssim \SI{100}{\nano\meter}$. To ensure that any likewise determined modulations of the phase solely arise from local variations of the in-plane magnetic induction in the sample, we restrict our considerations to small areas of the carefully cut lamella, for which a homogeneous thickness can be assumed and phase variations originating from the electrostatic potential of the sample can thus be neglected. 
Figures~\ref{Fig:LTEM_PhaseAnalysis}(a-c) show the resulting maps of the magnetic phase shift for one variant of stripe patterns acquired in out-of-plane magnetic fields, $\mu_0 H_\mathrm{ext}$, of 0, 150, and \SI{300}{mT}. At zero magnetic field [cf.~Fig.~\ref{Fig:LTEM_PhaseAnalysis}(a)] the phase pattern consists of stripes of alternating bright and dark contrast. 
As the gradient contrast represents strength and sign of $\Phi$, according to Eq.~\eqref{Eqn_TIE}, in-plane magnetic induction occurs solely at the interfaces between stripes of alternating contrast, since only here, the gradient is non-zero. Also, the direction of this induction is opposite for changes of contrasts at the stripe borders from bright to dark and from dark to bright pointing along the positive and negative $x$ direction, respectively. This shows that the DWs bordering the stripe domains are $180^\circ$ ($\pi$) Bloch walls. 
It is apparent from the phase patterns that with increasing field, the dark stripes widen on the expense of the bright ones. Since with increasing field, the sample is successively magnetized along the field direction, the dark stripes can be identified as domains with a (constant) out-of-plane magnetization in positive $z$ direction. At sufficiently high magnetic fields, the width of the bright domains magnetized opposite to field direction, $d_1$, shrinks to just the DW width, $w$. At this point, each two  neighboring $180^\circ$ ($\pi$) Bloch walls combine to form a single $360^\circ$ ($2\pi$) DW. 

These conclusions that are solely drawn from the experimentally acquired LTEM images enable us to deduce the underlying spin textures and their modulation along the direction of the contrast modulation, referred to as propagation direction. In Figs.~\ref{Fig:LTEM_PhaseAnalysis}(g-i) the resulting textures are shown for fields of 0, 150, and \SI{300}{mT}. At zero magnetic field [cf.~Fig.~\ref{Fig:LTEM_PhaseAnalysis}(g)], compensated up and down magnetized domains of roughly equal widths ($d_1 \simeq d_2$) are separated by $\pi$ Bloch walls of width $w$, thereby forming a stripe domain pattern. 
The DWs are of right-handed chirality as defined with respect to the propagation direction along the positive $x$ axis.
The resulting magnetic texture compares to that of neighboring out-of-plane magnetized stripe domains. 
In contrast to conventional stripe domains, however, these stripe domains form a {\em periodic lattice} of well defined periodicity. Accordingly, and in order to account for the connecting $\pi$ DWs, we refer to this magnetic ground state as $\pi$-CSL.

Upon increasing the field, domains magnetized in field direction start to grow at the expense of those magnetized opposite to the field [$d_2 > d_1$, see Fig.~\ref{Fig:LTEM_PhaseAnalysis}(h)], while $w$, which is fixed by the exchange constants (see below), remains largely constant. 
At $\mu_0 H_\mathrm{ext} = \SI{300}{mT}$, only domains magnetized along the field remain ($d_2 > d_1 = 0$), and pairs of neighboring $180^\circ$ ($\pi$) Bloch walls coalesce to form solitonic $360^\circ$ ($2\pi$) DWs of width $2w$ [cf.~Fig.~\ref{Fig:LTEM_PhaseAnalysis}(i)]. This spin texture resembles closely that of a conventional CSL.

In order to validate the likewise deduced 1D spin textures, they were extended along the $x$ direction to form 2D stripes, based on which both, phase images and maps of the in-plane magnetic induction, $B_y$, were then calculated using the \texttt{Ubermag} sub-package \texttt{mag2exp} \cite{beg2022} following the scheme of Beleggia and Zhu \cite{Bellagia2003}. 
As can be seen from a comparison of the images in Figs.~\ref{Fig:LTEM_PhaseAnalysis}(a-f) with those in Fig.~\ref{Fig:LTEM_PhaseAnalysis}(j-o), the experimental and the model-derived phase images and induction patterns are in very good agreement thereby confirming the deduced texture models.

\begin{figure}[t]
    \includegraphics[width=0.25\textwidth]{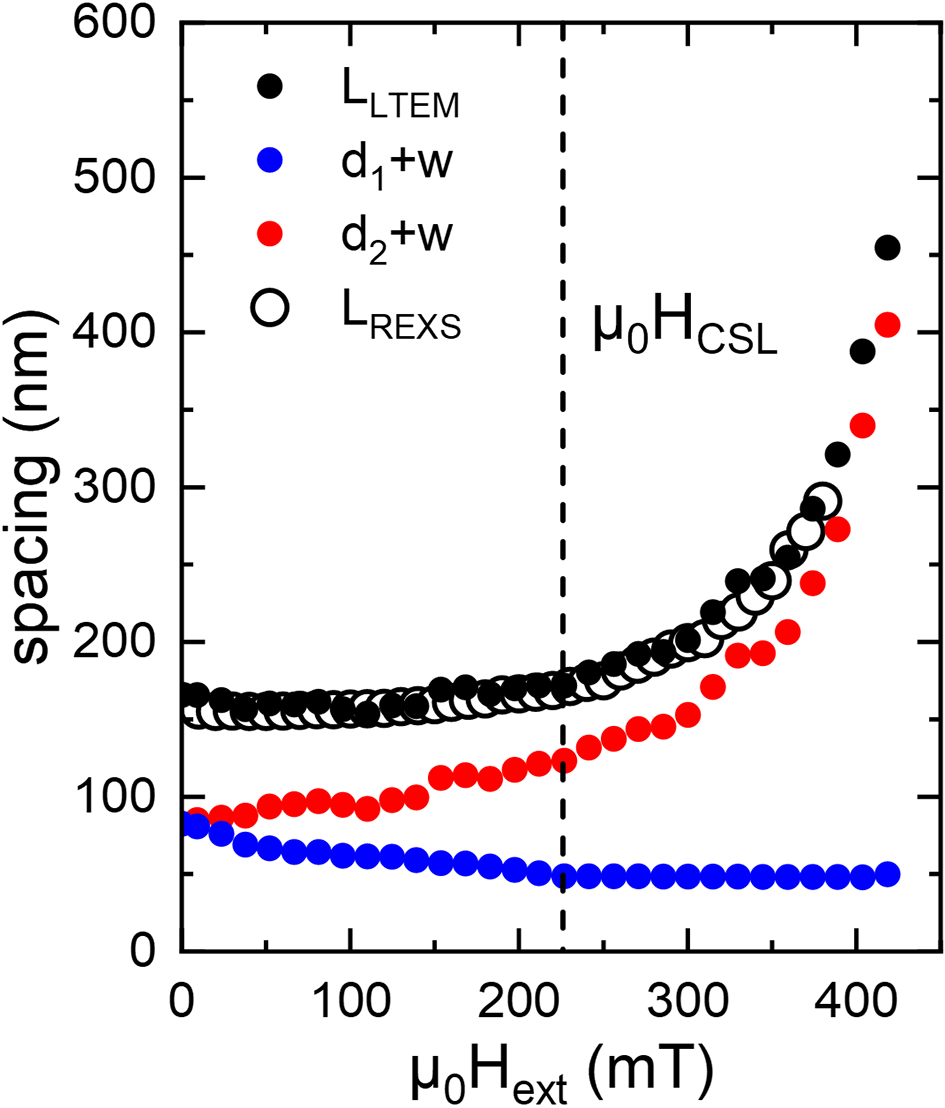}
    \caption{Measured values of the stripe patterns (solid circles) obtained from an analysis of the LTEM-derived phase patterns shown in Fig.~\ref{Fig:LTEM_PhaseAnalysis} as a function of the applied magnetic field, $\mu_0 H_\textrm{ext}$. Plotted are the periodic length, $L_\mathrm{LTEM}$, width of bright (dark) stripes, $d_1$ ($d_2$) including the domain wall width, $w$. 
    The periodic lengths obtained from the complementary REXS experiments, $L_\mathrm{REXS}$ (open circles) is also included for comparison. The vertical dashed line indicates the transition from a $\pi$-CSL to a conventional CSL. See text for details.
    }
    \label{Fig:Results_LTEM_PhaseAnalysis}
\end{figure}

Aside from the $d_2/d_1$ disproportion, the applied field also increases the modulation length in a characteristic way. Figure~\ref{Fig:Results_LTEM_PhaseAnalysis} shows this modulation length as determined from both  LTEM, $L_\mathrm{LTEM}$, and  REXS, $L_\mathrm{REXS}$, together with the LTEM-derived widths of the stripe domains, $d_1 + w$ and $d_2 + w$, as function of the applied field. The modulation lengths determined independently in two different experiments on the very same sample coincide perfectly. Both, $L_\mathrm{LTEM}$ and $L_\mathrm{REXS}$ grow monotonically with increasing field, and diverge upon approaching magnetic saturation in the ferromagnetically polarized state at $H_\mathrm{C}$. 
The dashed vertical line in Fig.~\ref{Fig:Results_LTEM_PhaseAnalysis} indicates the field, where $d_1$ vanishes and each two neighboring 180° ($\pi$) Bloch walls coalesce to form a $2\pi$ soliton each.
As this marks the transition from a $\pi$-CSL to a conventional CSL, we refer to this field as $H_\mathrm{CSL}$ with $\mu_0 H_\mathrm{CSL}\simeq \SI{226}{mT}$. 
Above $H_\mathrm{CSL}$, we find $d_1+w\approx w = 46\,$nm. Hence,  the characteristic width of a 2$\pi$ soliton in Mn$_{1.4}$PtSn is $2w=92$\,nm.

\subsection{The reciprocal space signature of the CSL from REXS experiments}

The accompanying REXS experiments are summarized in Fig.\ \ref{Fig:CSL_REXS_Analysis}. 
Fig.\ \ref{Fig:CSL_REXS_Analysis}(a) shows the field dependence of the magnitude of the scattering vector as determined from the $\rm 1^{st}$ order magnetic Bragg reflections (cf.\ Fig.\ \ref{Fig:LTEM_REXS_Exp}) normalized to their values at $H_\mathrm{ext} = 0$ along the crystallographic $a$ and $b$ axes $q(H_\mathrm{ext})_{1,a}$ and $q(H_\mathrm{ext})_{1,b}$, i.e., for both observed domain variants. 
We refer to this vector as the propagation vector of the chiral spin modulation. 
As revealed in Fig.\ \ref{Fig:Results_LTEM_PhaseAnalysis}, the increase in periodic length (i.e., the inverse of the magnitude of the propagation vector) is identical for both domains.
The steepest decrease in $q(H_\mathrm{ext})$ and increase in $L(H_\mathrm{ext})$ occurs at fields above $\mu_0 H_\textrm{ext} = \mu_0 H_\textrm{CSL}$, i.e., in the stability range of the conventional CSL.

The REXS experiments also support the finding from our LTEM analysis that the magnetic ground state in \mps\ is a $\pi$-CSL rather than the helical phase. 
Figure \ref{Fig:CSL_REXS_Analysis}(b) shows the field dependence of the normalized intensities $I_{{i,a/b}}(H_\text{ext})/I_{{1,a/b}}(0)$ of the $\rm 1^{st}$, $\rm 2^{nd}$, and $\rm 3^{rd}$ harmonic of the propagation vector as determined from the $\rm 1^{st}$, $\rm 2^{nd}$, and $\rm 3^{rd}$ order magnetic Bragg reflections. 
While for the sinusoidal spin modulation of a helical phase, only the fundamental spatial frequency (i.e., the $\rm 1^{st}$ order reflection) is expected to occur with a monotonically decreasing intensity, also higher harmonics are observed that even exhibit non-monotonic field dependencies. 
At low fields, the REXS pattern contains only the $\rm 1^{st}$ and $\rm 3^{rd}$ harmonics, while the $\rm 2^{nd}$ harmonic sets only in above around $\SI{30}{mT}$. 
Its intensity increases and reaches its maximum at \SI{260}{mT}.
Specifically, the behavior of the $\rm 3^{rd}$ harmonic, which is found to vanish at intermediate fields around $\SI{150}{mT}$ and then to rise again until all intensities are suppressed upon reaching full saturation in the ferromagnetic state, contradicts a continuous sinusoidal modulation. 
It is rather typical for a more discontinuous Heaviside or boxcar type of modulation with unequally long up and down segments. 
Similar non-sinusoidal modulations and domain wall sharpening effects have been observed in multilayered magnetic systems, where higher harmonics and soliton coalescence play a role~\cite{Shen2022}.
Then, multiples of the $n^\textrm{th}$ harmonic are suppressed when the ratio of up to down segments is $\frac{1}{n}/\frac{(n-1)}{n}$. 
For the $\rm 3^{rd}$ harmonic, this would relate to a ratio of the widths of up and down domains of $d_2/d_1 = 1/2$, which according to the LTEM analyses is fulfilled below roughly $\SI{140}{mT}$ (cf.~Fig.~\ref{Fig:LTEM_PhaseAnalysis}). A similar behavior is observed, e.g., in stripe domains of ferromagnetic multilayers~\cite{Hellwig2003}.

\section{Theoretical validation and simulations}

\subsection{The double Sine-Gordon model}

In the following, we will show with the aid of theoretical considerations that the observed discrepancies with respect to both a helical ground state and a classical CSL are due to a delicate balance of the (competing) magnetic anisotropies in the sample.

The non-centrosymmetric half-Heusler compound \mps can be modeled within the $D_{2d}$ Hamiltonian
\begin{equation}
\label{eq.D2d-Hamilton}
    \mathcal{H} = \int d\vect{r}\left[ \epsilon_\mathrm{ex} + \epsilon_{\mathrm{DMI}} + \epsilon_\mathrm{Zeeman} + \epsilon_\mathrm{MCA}  + \epsilon_\mathrm{MS} \right],
\end{equation}
with
\begin{equation}
\label{eq.engergy-densities}
    \begin{array}{lll}
        \epsilon_\mathrm{ex} & = & A(\vect{\nabla} \vect{m})^2,  \\
        \epsilon_{\mathrm{DMI}} & = & D(m_z \partial_x m_y - m_y \partial_x m_z \\
        & & \hspace{52 pt} - \hspace{2 pt}m_x \partial_y m_z+ m_z \partial_y m_x),\\
        \epsilon_{\mathrm{Zeeman}} & = & - \mu_0 M_s \vect{m}\cdot\vect{H}_{\mathrm{ext}}, \\
        \epsilon_{\mathrm{MCA}} & = & - K (\vect{m}\cdot\vect{\hat{n}})^2, \\
        \epsilon_{\mathrm{MS}} & = & - \frac{1}{2}\mu_0 M_s \vect{m}\cdot\vect{H}_{\mathrm{MS}},
    \end{array}
\end{equation}

\begin{figure}[t]
    \includegraphics[width=0.46\textwidth]{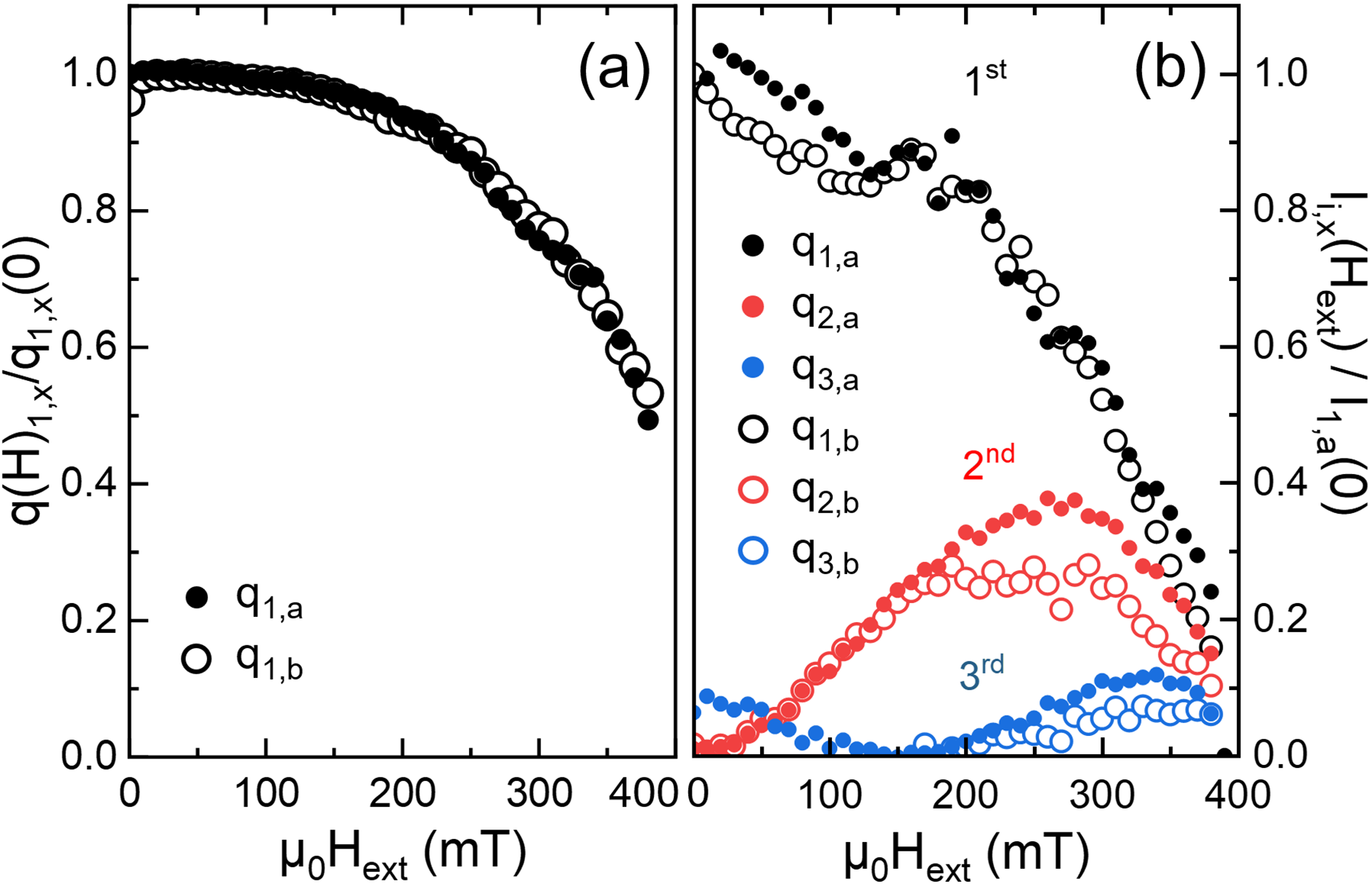}
    \caption{Field dependence of (a) the normalized magnitude of the propagation vector and (b) the intensities of its $\mathrm{1^{st
    }}$, $\mathrm{2^{nd}}$, and $\mathrm{3^{rd}}$ harmonics as determined from the $\mathrm{1^{st}}$, $\mathrm{2^{nd}}$, and $\mathrm{3^{rd}}$ order diffraction spots (cf.~Fig.~\ref{Fig:LTEM_REXS_Exp}).
    }
    \label{Fig:CSL_REXS_Analysis}
\end{figure}

\noindent
representing the energy densities due to the symmetric exchange interaction, the ${D_{2d}}$ symmetry DMI~\cite{Dzyaloshinkskii1964,Bogdanov1994,Bogdanov1999}, the Zeeman interaction, the uniaxial magnetocrystalline anisotropy (MCA), and the magnetostatic (MS) interaction, respectively. 
Interactions are taken into account up to the second order in spin-orbit coupling.
$A$ is the ferromagnetic exchange stiffness, $D$ the strength of the DMI, $\vect{m}$ the unit vector field describing the magnetization direction, $\mu_\mathrm{0}$ the vacuum magnetic permeability, $M_s$ the saturation magnetization, $\vect{H}_\mathrm{ext}$ the strength of the external magnetic field, $K$ the uniaxial magnetocrystalline anisotropy constant, $\vect{\hat{n}}$ the unit vector defining the anisotropy axis, and $\vect{H}_\mathrm{MS}$ the magnetostatic field~\cite{Kravchuk2023}.

The $D_{2d}$ symmetry DMI favors opposite-handed helical twisting along the two perpendicular directions $x$ and $y$, which manifests itself in the different LTEM contrast of the stripes in domains with perpendicular propagation directions in Figs.~\ref{Fig:LTEM_REXS_Exp}(c,e). 
For \mps, the crystallographic $a$, $b$, and $c$ axes correspond to $x$, $y$, and $z$ directions in Eq.\ \eqref{eq.engergy-densities}, respectively, where $z$ is oriented along the film normal.

Note that the nature of the spin modulation in \mps\ differs from that of a classical CSL. 
Conventional CSLs in monoaxial helimagnets have an easy \emph{plane} (uniaxial) magnetic anisotropy. In contrast, \mps\ exhibits an easy {\em axis} anisotropy.
While in the former, the magnetic moments $\vect{m}$ rotate in a plane perpendicular to the anisotropy axis $\vect{\hat{n}}$ and the uniaxial anisotropy term $\epsilon_{\mathrm{MCA}}$ in Eqs.\ \eqref{eq.D2d-Hamilton} and \eqref{eq.engergy-densities} becomes zero, in \mps, $\vect{\hat{n}}$ points along the easy axis of magnetization and thus results in finite contributions of $\epsilon_{\mathrm{MCA}}$ to the total energy density of the system.

A thorough consideration of long-range magnetostatic interactions requires to take into account the details of the sample geometry and orientation as well as that of the spin textures therein, and will thus necessitate micromagnetic simulations. 
However, also analytical predictions can be derived, if the magnetostatic interaction is approximated by assuming a homogeneously magnetized sample with a magnetization that is identical to that of the more complex real magnetic texture.
To this end, a constant $K_\mathrm{MS}$ representing the magnetostatic interactions is added to the magnetocrystalline anisotropy to define a modified effective magnetic anisotropy, $\Tilde{K}= K_\mathrm{MCA} + K_{\mathrm{MS}}$.

To describe the experimentally observed single-domain helical spirals, where the external magnetic field lies along the $z$-direction, i.e., $\vect{H}_\mathrm{ext} = H_\mathrm{ext} \vect{\hat{z}}$, Eq.~\eqref{eq.D2d-Hamilton} can then be reduced to an effective one-dimensional model along a main symmetry axis, e.g., the $\hat{x}$ direction, where the model along $\hat{y}$ differs by the sign of the DMI. 

Inspired by the results of our LTEM experiments, for the magnetization vector, we make the ansatz 
\begin{equation}
\label{eq.ansatzhelix}
    \vect{m}=(0,-\sin\phi,\cos\phi),
\end{equation}
which leads to an effective 1D model for the angle $\phi$
\begin{equation}
\begin{split}
\label{eq.1dmodel}
     \mathcal{H}(\phi) & = \\
     S\int dx & \left[ A \partial^2_x \phi - D \partial_x \phi - \mu_0 M_s H_\mathrm{ext} \cos\phi - \Tilde{K} \cos^2\phi \right],
\end{split}
\end{equation}
where $S$ is the area of the sample in the $yz$-plane. The functional form of the profile function $\phi=\phi(x)$ depends on the parameters of Eq.~\eqref{eq.1dmodel}, which determine whether one obtains a distorted helical spiral with a propagation vector along the $x$-direction, or a single soliton. 

To obtain dimensionless variables, we rescale lengths by the propagation vector in the absence of anisotropy and magnetic field, $Q$, the magnetic field by the critical field , $H^\ast$, and the anisotropy by its critical value at zero field, $K^\ast$:
\begin{equation}
\label{eq.normalizations}
    \xi = x Q, \quad h=\frac{H_\mathrm{ext}}{H^\ast} \quad \kappa = \frac{\Tilde{K}}{K^\ast},
\end{equation}
where $Q$, $H^\ast$ and $K^\ast$ are defined as
\begin{equation}
         Q = \frac{D}{2A}, \quad H^\ast=\frac{D^2}{2\mu_0M_sA}, \quad K^\ast = \frac{D^2}{2A},
\end{equation}
respectively. 
Expressing Eq.~\eqref{eq.1dmodel} in the above introduced naturally rescaled units yields
\begin{equation}
\label{eq.1d_model_phi}
    \frac{\mathcal{H}(\phi)}{SE_0}= \int d \xi \left[ \frac{1}{2} \phi'^2 - \phi' - h \cos \phi - \kappa \cos^2 \phi\right],
\end{equation}
where $E_0 = D $ is the unit of energy,  $\phi=\phi(\xi)$, and prime denotes the derivative with respect to the dimensionless length $\xi$. 

Minimizing the energy functional, Eq.~\eqref{eq.1d_model_phi} with respect to the helical spiral's profile $\phi(\xi)$
yields the \emph{double sine-Gordon equation} 
\begin{equation}
\label{eq.phi0}
    \phi''(\xi)  = h \sin \phi(\xi) +\kappa \sin2\phi(\xi),
\end{equation}
which can be integrated to
\begin{equation}
\label{eq.phi1}
     \frac{1}{2}(\phi'(\xi))^2  = - h \cos \phi(\xi)  + \kappa \sin^2 \phi(\xi) + c.
\end{equation}
Here, $c$ is an integration constant, which is determined by minimizing the energy per unit length~\cite{Kravchuk2023}. 
Note that Eq.~\eqref{eq.phi0} is solved by the ferromagnetic state, i.e., the trivial homogeneous solutions with $\phi=N\pi$, where $N\in\mathbb{Z}$.

\begin{figure}[t]
    \includegraphics[width=0.35\textwidth]{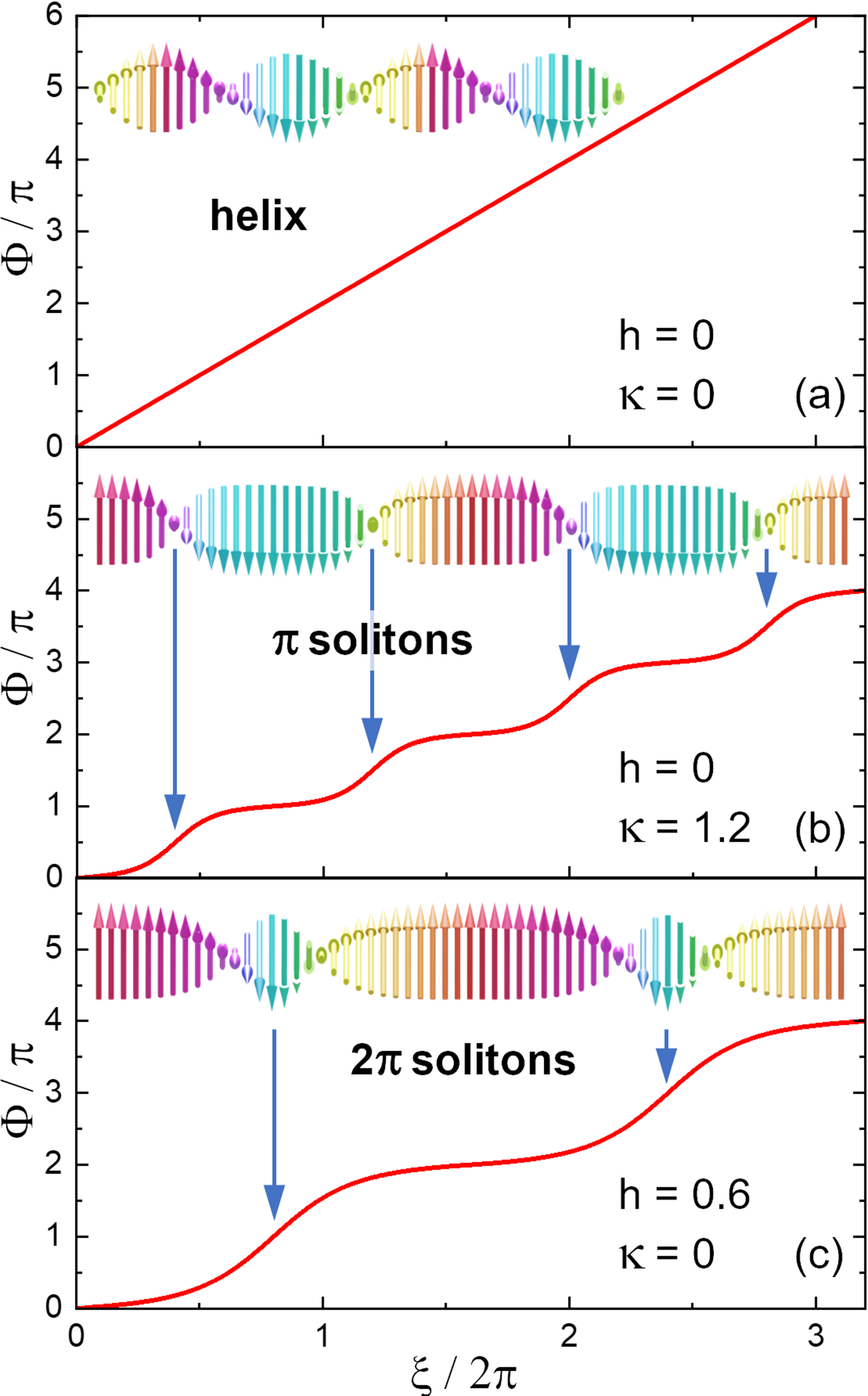}
    \caption{Effect of magnetic anisotropy and field on the chiral structure as derived from 1D sine-Gordon models: Phase $\Phi$ of the coplanar rotation as a function of the dimensionless length $\xi$ along the direction of propagation. 
    (a) Helical phase with neither field nor anisotropy. 
    (b) $\pi$-CSL in zero field at finite anisotropy. 
    (c) Conventional CSL with $2\pi$ solitons at finite fields.}
    \label{Fig:single_SG_solutions}
\end{figure}

For zero magnetic field ($h = 0$) or zero magnetic anisotropy ($\kappa = 0$), Eq.~\eqref{eq.phi0} can be solved analytically, as these borderline cases effectively reduce the double sine-Gordon notation to single sine-Gordon models each. Noteworthy, and as already conceivable from a comparison of the two terms in Eq.~\eqref{eq.phi0}, the modulation length varies twice as strongly with the field at vanishing anisotropy than with the anisotropy at zero field, respectively.

\subsection{Magnetic phase diagram and the impact of anisotropies}

Figure~\ref{Fig:single_SG_solutions} shows the solutions of the according single sine-Gordon equation for three different scenarios. 
At both zero field and zero anisotropy, one obtains a purely sinusoidal modulation that represents the helical ground state [$h = 0$, $\kappa = 0$, Fig.~\ref{Fig:single_SG_solutions}(a)]. 
Upon introducing a magnetic anisotropy, the ground state at zero field changes into a $\pi$-CSL with equally wide up and down domains [$h = 0$, $\kappa = 1.2$, Fig.~\ref{Fig:single_SG_solutions}(b)], while applying a magnetic field drives the system into a classical CSL [$h = 0.6$, $\kappa = 0$, Fig.~\ref{Fig:single_SG_solutions}(c)].

\begin{figure}[t]
    \includegraphics[width=0.48\textwidth]{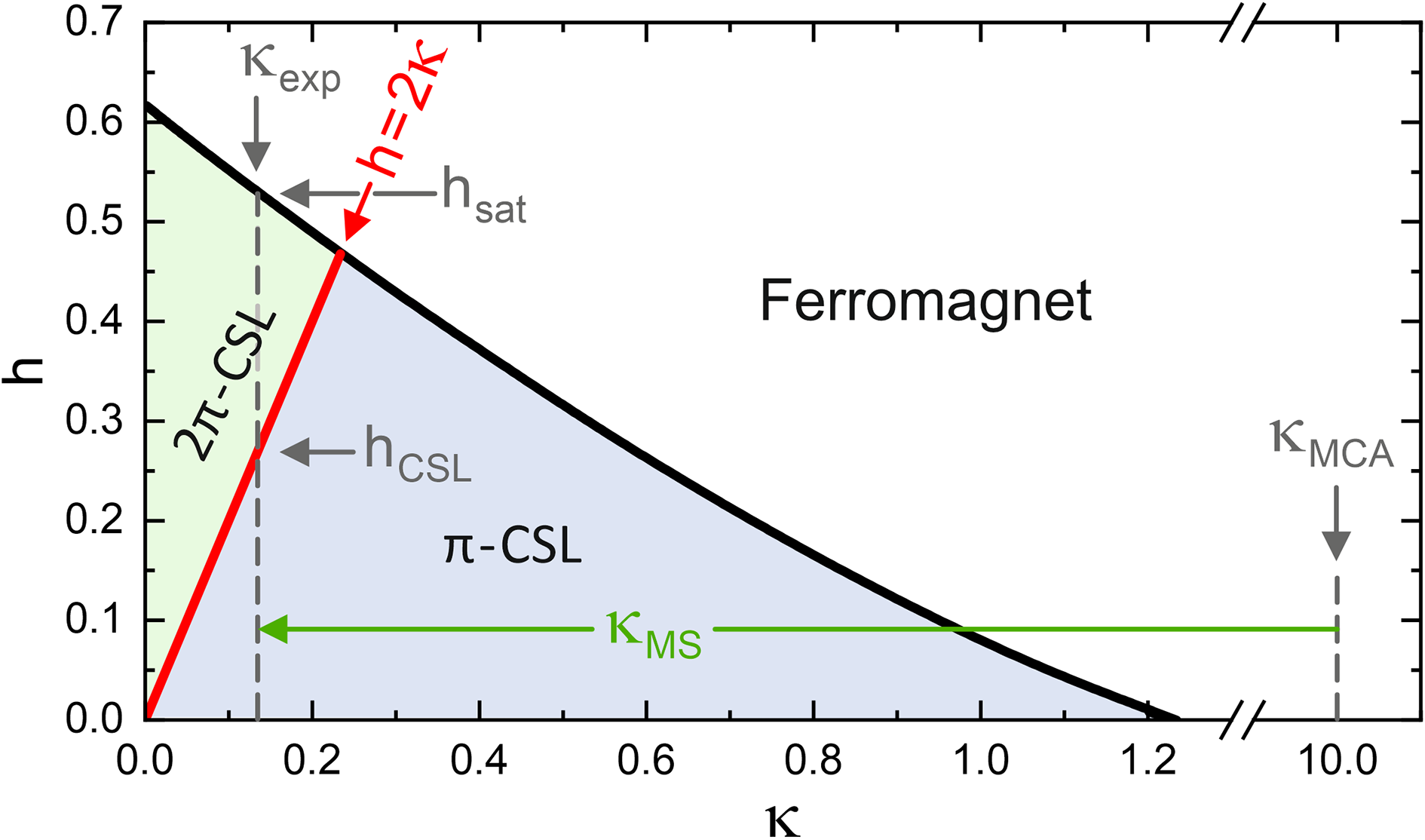}
    \caption{Magnetic phase diagram as a function of the normalized magnetic field $h$ and the normalized magnetic anisotropy $\kappa$ as derived from the double sine-Gordon model [cf.~Eq.~\eqref{eq.phi1}]. 
    The red line separates the stability region of the $\pi$-CSL (green region) from that of the $2\pi$-CSL (blue)~\cite{condat1983}, while the black line indicates the transition into the saturated ferromagnetic phase~\cite{Ross2021}. 
    } 
    \label{Fig:single_SG_PhasDiag}
\end{figure}

\begin{figure}[h]
    \includegraphics[width=0.45\textwidth]{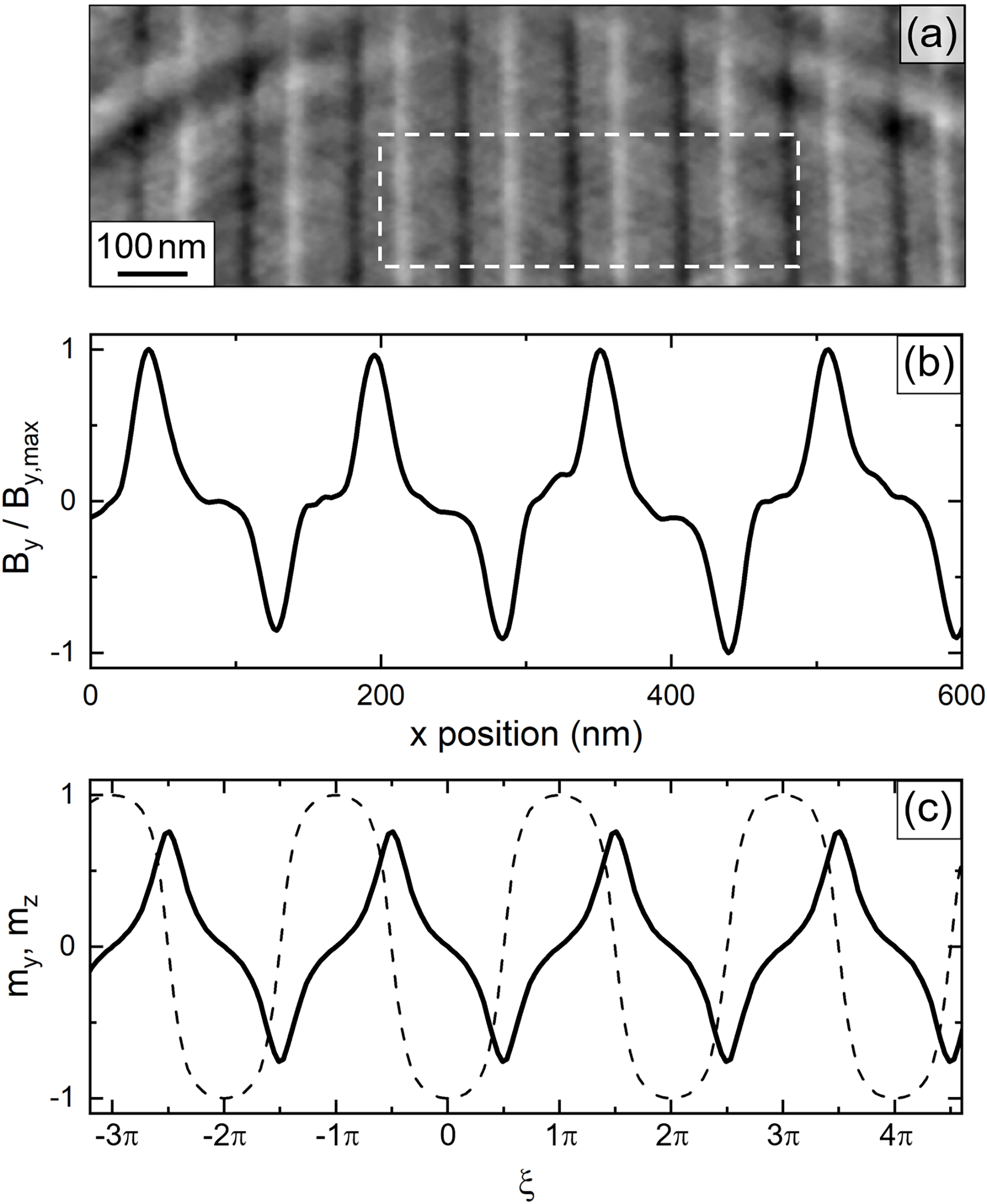}
    \caption{(a) Map of the $y$ component of the in-plane magnetic induction, $B_{y}$, derived from the experimental LTEM image acquired at zero magnetic field [cf.~Fig.~\ref{Fig:LTEM_PhaseAnalysis}(d)]. 
    (b) Averaged and normalized profile of the $y$ component of the in-plane magnetic induction, $B_{y}(x)$, as derived from the intensity variation in the area bordered by the dashed line in (a). 
    (c) Corresponding profiles of the $y$ (solid line) and $z$ components (dashed line) of the magnetization along the dimensionless length $\xi$ in the double sine-Gordon equation [cf.~Eq.\ \eqref{eq.phi0}] as obtained from micromagnetic simulations.
    }
    \label{Fig:Micromagnetics_dSG_role_of_anisotropy}
\end{figure}

The phase diagram in Fig.\ \ref{Fig:single_SG_PhasDiag} represents those regions in the ($h$, $\kappa$) parameter space, where according to the double sine-Gordon model, the 2$\pi$-CSL, the $\pi$-CSL, and the saturated ferromagnet are stable. The red line along the condition $h = 2\kappa$ separates the stability regions of the two different CSLs \cite{condat1983,Ross2021}: 
In the magnetic field-dominated region $h > 2\kappa$ (green area), the 2$\pi$-CSL forms, while in the anisotropy-dominated region $h < 2\kappa$ (blue), the $\pi$-CSL resides. 
The phase boundary between the two spiral phases on the one side and the ferromagnetic phase on the other can be derived analytically from the constraint that at the transition to the ferromagnetic state, the energy of the $2\pi$ soliton becomes zero for an anisotropy dependent critical field $h_c(\kappa)$ \cite{condat1983,Ross2021} that is defined by the relationship
\begin{equation}
\label{eq.phase_boundary}
    2\pi = 4\sqrt{h_c+2\kappa}+ \frac{4h_c}{\sqrt{2\kappa}} \, \mathrm{arcsinh}{\left(\sqrt{\frac{2\kappa}{h_c}} \right)}.
\end{equation}

The resulting variation of $h_c$ with $\kappa$ is plotted as a solid black line in Fig.\ \ref{Fig:single_SG_PhasDiag}. 
The normalized magnetocrystalline anisotropy, $\kappa_\mathrm{MCA}$ calculated from $K_\mathrm{MCA}$ = \SI{1e5}{Jm^{-3}} as reported by L.\ Peng et al.~\cite{Peng2020}, is also indicated in the diagram by the gray dashed line at the right. 
As can be seen, if the effective magnetic anisotropy would solely be due to the MCA, no other but the ferromagnetic phase should occur in the material. 
Our experiments, however, show that \mps\ exhibits a $\pi$-CSL at zero field that transforms into a 2$\pi$-CSL at $H_\mathrm{CSL}$, which is roughly half the saturation field $H_\mathrm{sat}$. 
This observation enables us to identify the effective magnetic anisotropy $\kappa_\mathrm{exp}$ prevailing in the sample, indicated by the gray dashed line on the left in Fig.\ \ref{Fig:single_SG_PhasDiag}. 
Since the likewise determined effective anisotropy is much smaller than the MCA, an additional contribution from the shape anisotropy, $\kappa_{\mathrm{MS}}$ of almost the same magnitude as $\kappa_\mathrm{MCA}$, is needed to (partially) compensate for the latter (cf.~green line in Fig.~\ref{Fig:single_SG_PhasDiag}). This qualitative conclusion is confirmed by frequent experimental observations of a distinct thickness dependence of the magnetic textures and phases occurring in \mps\ samples of different geometry~\cite{ZunigaCespedes2021,Winter2022,Karube2022,Ma2020}.
With increasing sample thickness, the shape anisotropy (and thus $K_\mathrm{MS}$) decreases, which leads to an increase of the periodic length of the soliton lattice and the successive stabilization of the ferromagnetically saturated state over the CSL.

\begin{figure}[t]
    \includegraphics[width=0.45\textwidth]{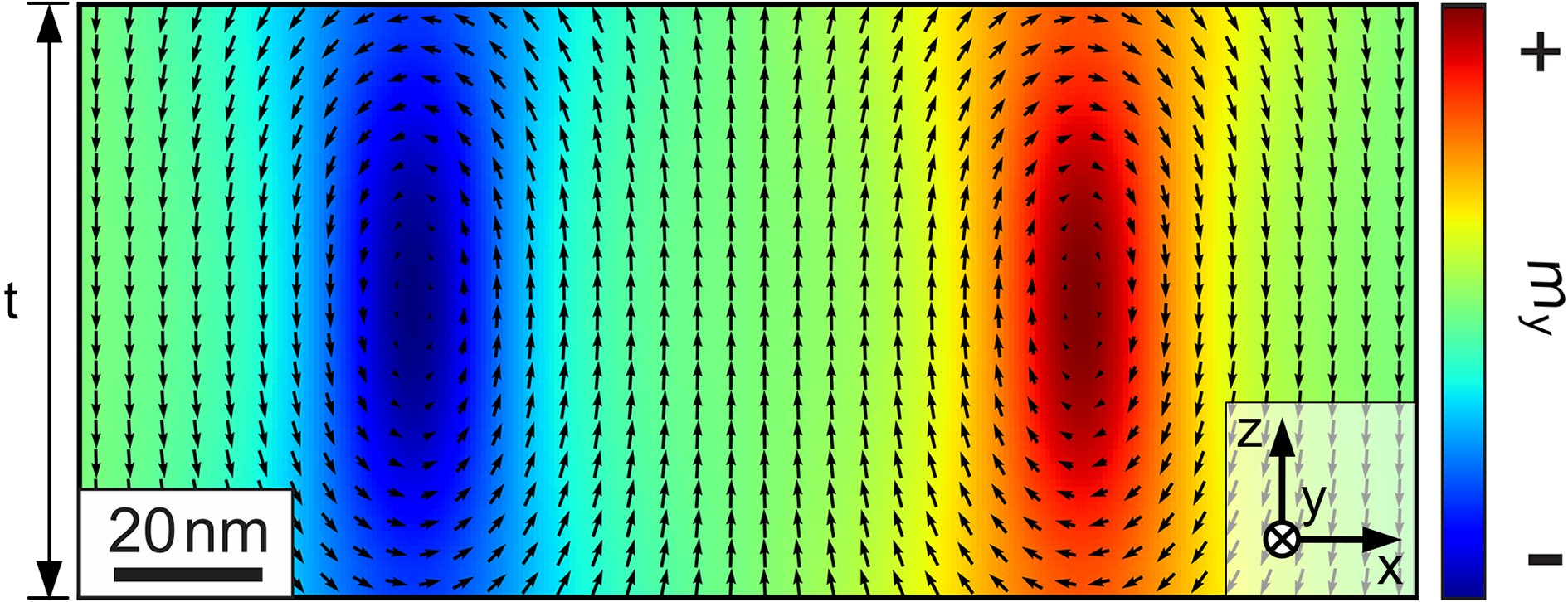}
    \caption{Color map of the $y$ component of the in-plane magnetization, $m_y(x,z)$, for a cross-section through the $xz$ plane of the sample as obtained from micromagnetic simulations. 
    Arrows indicate the (projected) length and orientation of the magnetization vectors in this cross-section, and $t$ denotes the sample thickness.
    }
    \label{Fig:Micromagnetics_my_profile}
\end{figure}

\subsection{Micromagnetic simulations}

Unlike in these borderline cases, for a more comprehensive and realistic scenario of a material with $D_{2d}$ symmetry and finite easy axis magnetic anisotropy in the presence of an arbitrarily chosen magnetic field, the double sine-Gordon model [Eq.\ \eqref{eq.phi1}] can no longer be solved analytically. For these general cases, the spin textures were rather derived from micromagnetic simulations.
We have therefore conducted such simulations based on the Hamiltonian described in Eq.\ \eqref{eq.D2d-Hamilton} for a 1D modulation (arbitrarily chosen along $x$) to investigate the experimentally observed textures in the stripe domains. 
Computational details of the simulations are described in the Methods section below.
In order to verify that the results of these simulations are consistent with our experimental observations, we compare in Fig.~\ref{Fig:Micromagnetics_dSG_role_of_anisotropy} the profile of the experimentally measured in-plane magnetic induction at zero magnetic field with the simulated magnetization profile. 
As can be seen from the solid lines in Figs.~\ref{Fig:Micromagnetics_dSG_role_of_anisotropy}(b,c), the normalized profiles of the experimentally determined $y$ component of the in-plane induction, $B_y / B_y^\mathrm{max}$, and the magnetization, $m_y / |m|$, are in very good qualitative agreement.
The observation that unlike for the $z$ component, $m_z$ [dashed line in Fig.~\ref{Fig:Micromagnetics_dSG_role_of_anisotropy}(c)], the maxima in $m_y(\xi)$ never reach the magnitude of the full magnetic moment, $|m|$, (i.e., $\textrm{max}(B_y / B_y^\mathrm{max}) < 1$) is due to the fact that identical to in the experimental situation, $m_y$ is averaged along $z$ over the total sample thickness. 
The mapping of $m_y$ in the cross-sectional $xz$ plane of the simulated sample in Fig.~\ref{Fig:Micromagnetics_my_profile} reveals the dipolar interaction-related occurrence of Néel-type surface states (often referred to as chiral surface twists \cite{Leonov2016,Schneider2018, Wolf2022}) where the magnetic moments are partially pointing along the $x$ direction. 
As a consequence, and unlike in the vertical center of the sample, averaging $m_y$ over the thickness results in a value that never reaches the magnitude of the full magnetic moment. 
In the experimental LTEM images, Néel-type textures remain ``invisible'' in out-of-plane imaging geometry, and since the intensity of the magnetic contrast in the experimental profile is normalized to its experimental maximum, $I/I_\mathrm{max}$ does reach unity in its maxima in Fig.\ \ref{Fig:Micromagnetics_dSG_role_of_anisotropy}(b). 

\begin{figure}[t]
    \includegraphics[width=0.48\textwidth]{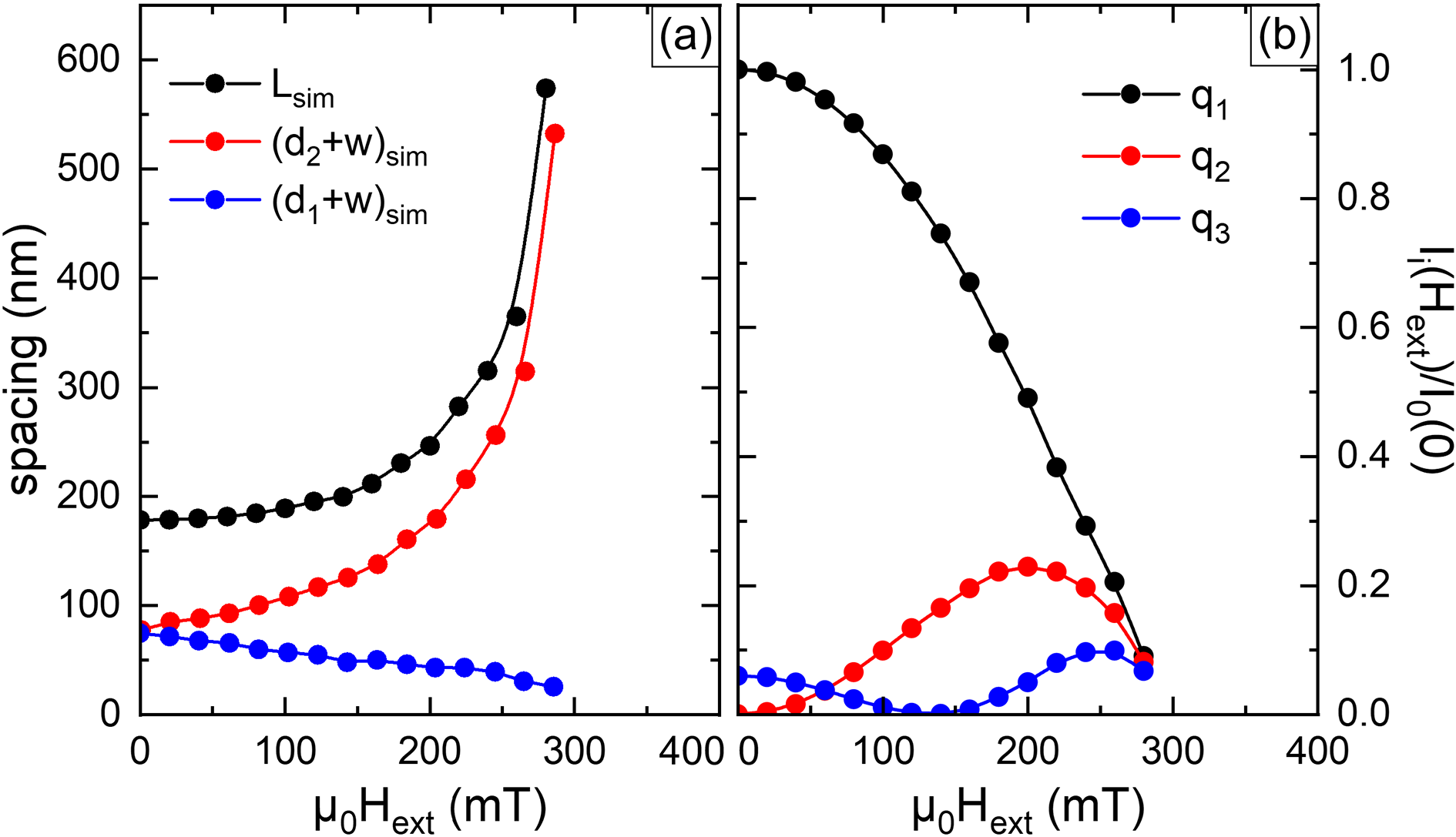}
    \caption{Micromagnetic simulations of the spin textures in \mps\ at finite magnetic field and anisotropy. (a) 
    Modulation length, $L_\mathrm{sim}$ (black dots) and domain widths, $(d_\mathrm{1,2} + w)_\mathrm{sim}$, (red and blue dots). (b) Intensity of the first three harmonics of the propagation vector $q_1$, $q_2$, and $q_3$ as a function of the magnetic field determined from the spin textures resulting from micromagnetic simulations of the double sine-Gordon model with identical parameters as in (a).
    }
    \label{Fig:dSG_micromagnetics}
\end{figure}

In a second step we investigate, if and to which extend the micromagnetic simulations also allow to describe the magnetic field-dependence of the experimental LTEM and REXS patters. 
Figure~\ref{Fig:dSG_micromagnetics}(a) shows the field dependence of the periodic length, $L_\textrm{sim}$, and the widths of domains with moments aligned parallel ($d_2 +w$) and antiparallel to the external field ($d_1 +w$). 
The simulation are found to not only nicely reproduce the experimentally determined geometrical measures of the CSL stripe patterns presented in Fig.~\ref{Fig:Results_LTEM_PhaseAnalysis}. 
Also the field-dependence of the $\mathrm{2^{nd}}$ and $\mathrm{3^{rd}}$ harmonics of the propagation vector, specifically the finite value of the intensity of the $\mathrm{3^{rd}}$ harmonic at zero field and its non-monotonic course  are reproduced [cf.~Figs.~\ref{Fig:CSL_REXS_Analysis} and \ref{Fig:dSG_micromagnetics}(b)].

Intriguingly, the results of our micromagnetic simulations conducted within the framework of the \emph{double sine-Gordon} model are in excellent agreement with our experimental observations.
They confirm the conclusion drawn from the LTEM and REXS experiments that at low magnetic fields, the sample exhibits a $\pi$-CSL rather than a helical modulation in the ground state, which then gradually transforms to a 2$\pi$-CSL upon increasing the magnetic field.

\section{Conclusion}

We have explored in detail the magnetic textures in the Heusler compound \mps, utilizing LTEM and REXS for high-resolution real space magnetic imaging and complementary reciprocal space magnetic scattering on the very same samples and under identical conditions at room temperature. 
The experiments were supported by micromagnetic simulations and successive simulations of LTEM images and REXS patterns.

The basic magnetic motif in \mps, which defines both its ground state properties and its field-dependent variations, is found to be the CSL. 
Unlike previously discussed, the ground state is a periodic arrangement of magnetic (stripe) domains with alternating orientation along and perpendicular to the external field separated by $\pi$ domain walls. 
Due to its periodic nature and the field-induced control of the periodic length, we identify this state as a $\pi$-CSL. 
The LTEM and REXS experiments have consistently shown that with increasing out-of-plane magnetic field along the uniaxial magnetocrystalline easy $c$ axis the $\pi$-CSL eventually transforms into a ``classical'' $2\pi$-CSL at a critical field $\mu_0H_\textrm{CSL}$. 
Due to the anisotropic DMI, these CSLs occur in two variants of opposite chirality, the helical propagation vectors of which align with the two DMI vectors along the $a$ and $b$ directions of the crystal.

A double sine-Gordon model, which takes into account the impact of both an external magnetic field and the magnetocrystalline anisotropy, allowed us to deduce a magnetic phase diagram that reproduces the experimentally observed $\pi$-CSL, $2\pi$-CSL, and ferromagnetically saturated state. 
In the limiting cases of vanishing fields on the one hand and vanishing anisotropy on the other, the model reduces to two single sine-Gordon equations that reveal both the anisotropy-stabilized $\pi$-CSL as the magnetic ground state and its field-driven transformation into a $2\pi$-CSL.
Micromagnetic simulations based on the Hamiltonian of the double sine-Gordon model further allowed us to explore the effect of the combined impact of magnetic fields and anisotropy for $D_{2d}$ or $S_{\rm{4}}$ systems and tie in closely with our experimental observations in \mps.

Crucially, this implies that the (double) sine-Gordon equation can be used as a starting point to clarify magnetic phase diagrams and to interpret transport and bulk physical properties of these structural families, independent of the specific materials details. 
One important and timely application will be the distinction of anomalous and topological Hall effect contributions in nanoscopic samples, where magnetometry is rarely feasible and mandatory in-situ techniques have only recently been developed~\cite{Winter2022,Pohl2023,Thomas2025}. 
It is also conceivable that this concept will inspire intuitive explanations for the formation of complex non-coplanar 2D textures like bubble lattices and antiskyrmion lattices, which have also been observed in \mps.

\section*{Experimental Methods}
\small
The REXS experiments reported here were performed at beamline I10 at the Diamond Light Source (UK) using the portable octupole magnet system (POMS) end station~\cite{VANDERLAAN201495}. 
The large wavelength at the Mn $L_3$ edge (637\,eV, 19.5\,\AA) relative to the lattice spacing dictates that the experiment must be performed in transmission, i.e., with momentum transfers in the first Brillouin zone. 
Supplementary measurements that enabled the present insights were carried out at BL29-MaReS at ALBA (Spain). 
LTEM measurements were performed using a JEOL F-200 microscope. 
In the Lorentz mode of operation, the objective lens was used to apply magnetic fields along the axis of the microscope column, i.e., perpendicular to the sample plane. 
All measurements reported here were performed at room temperature.

\section*{Computational Methods}
\label{comp.methods}
Micromagnetic simulations were conducted using \texttt{Mumax$\mathrm{^3}$} \cite{vansteenkiste_design_2014}. 
We used the following values for the exchange stiffness, Dzyaloshinskii-Moriya interaction constant, saturation magnetization, and uniaxial anisotropy constant: $A$ = \SI{8e-12}{J m^{-1}}, $D$ = \SI{4e-4}{J m^{-2}}, $M_\mathrm{S}$ = \SI{4.5e5}{Am^{-1}}, and $K_\mathrm{MCA}$ = \SI{1e5}{J m^{-3}}, respectively~\cite{Peng2020}. 
Dipolar interactions were explicitly included in the simulations. The sample dimensions were set to $L_x \times2$ nm $\times$ $t$ (length $\times$ width $\times$ thickness along the $x$, $y$, and $z$ directions, respectively). Here, $L_x$ represents the modulation length along $x$ as determined from energy minimization. The width was limited to \SI{2}{nm}, since due to the anisotropic DMI, for modulations along $x$, no variation along $y$ was to be expected. In order to account for the fact that the determination of the lamella thickness in the FIB is easily overestimated due to the Pt protection layer, in the simulations, the thickness was adjusted to $t =$ \SI{80}{nm} so that the resulting periodic length at zero field matches periodic length $L_{\mathrm{LTEM}}$ determined from the LTEM experiments. The mesh size was set to 1$\times$1$\times$1 nm$^3$, and the simulations were carried out under periodic boundary conditions in the $xy$ plane and an open boundary condition along $z$.

\section*{Acknowledgments}
M.W.\ acknowledges support from the International Max Planck Research School for Chemistry and Physics of Quantum Materials (IMPRS-CPQM).
M.C.R.\ was supported through the Emmy Noether program of the German Research Foundation, DFG, (project no.\ 501391385). Work at TU Dresden was supported by the DFG through CRC 1143 and the Cluster of Excellence {ct.qmat} (EXC 2147, project no.\ 390858490).
K.E.S.\ acknowledges funding from the DFG through project no.\ 320163632 and through project no.\ 403233384 within SPP 2137.
S.S., D.P., and B.R.\ are grateful for funding from the DFG through SPP 2137, project no.\ 403503416.
D.P.\ acknowledges financial support from the DFG through project no.\ 504660779.
REXS was conducted using the Portable Octupole Magnet System on beamline I10 at the Diamond Light Source, UK, under proposal MM28882 and at BL29-MaReS at the ALBA Synchrotron (Spain), under proposal 20220672. 
Financial support by the EPSRC (EP/N032128/1) is gratefully acknowledged.

\section*{Author Contributions}
M.W., M.C.R., C.F.\ and B.R.\ devised the project. M.W.\ conducted the LTEM experiments with the help of D.P., S.S., and D.W. 
M.W., M.C.R, A.S.S., B.A., J.R.B., V.U., M.V., G.vdL., and Th.H.\ conducted the REXS experiments. A.P., M.A., and K.E.S.\ developed the theory and conducted micromagnetic simulations in close collaboration with M.W. 
M.W.\ analyzed all data. 
P.V.\, A.T.\, and M.W.\ prepared the samples with the support of T.H. 
M.W., B.R., M.C.R., Th.H., A.P., and M.A.\ wrote the manuscript. All authors contributed to the discussion and reviewed the manuscript.

\section*{Data availability}
The data are not publicly available. However, the data are available from the authors upon reasonable request.


%

\end{document}